\pdfoutput=1

\documentclass{article}
\usepackage{ismir,amsmath,cite,url}
\usepackage{graphicx}
\usepackage{color}

\newcommand\nj[1]{\textcolor{black}{#1}}
\newcommand\sh[1]{\textcolor{black}{#1}}
\newcommand\kg[1]{\textcolor{black}{#1}}

\newcommand\shrev[1]{\textcolor{black}{#1}}

\newcommand{\etal}{\textit{et al}.}

\title{Music Source Separation Using Stacked Hourglass Networks}

\multauthor
{Sungheon Park \hspace{1cm} Taehoon Kim \hspace{1cm} Kyogu Lee \hspace{1cm} Nojun Kwak}
  {Graduate School of Convergence Science and Technology, Seoul National University, Korea\\
 {\tt\small \{sungheonpark, kcjs55, kglee, nojunk\}@snu.ac.kr}
}
\def\authorname{Sungheon Park, Taehoon Kim, Kyogu Lee, Nojun Kwak}

\begin{document}

\maketitle
\begin{abstract}
In this paper, we propose a simple yet effective method for multiple music source separation using convolutional neural networks. Stacked hourglass network, which was originally designed for human pose estimation in natural images, is applied to a music source separation task. The network learns features \kg{from} a spectrogram image across multiple scales and generates masks for each music source. The estimated mask is refined as it passes over stacked hourglass modules. The proposed framework is able to separate multiple music sources using a single network. Experimental results on MIR-1K and DSD100 datasets validate that the proposed method \kg{achieves} competitive results \kg{comparable} to the state-of-the-art methods in multiple music source separation and singing voice separation tasks.
\end{abstract}
\section{Introduction}\label{sec:intro}

Music source separation is one of the fundamental research areas for music information retrieval. Separating singing voice or sounds of individual instruments from \kg{a mixture} has grabbed a lot of attention in recent years. The separated sources can be \kg{further} used for applications such as \kg{automatic music transcription, instrument identification, lyrics recognition, and so on.}

Recent improvements on deep neural networks (DNNs) \nj{have been blurring} \kg{the} boundaries \kg{between many application domains, including}  computer vision and audio signal processing. Due to its end-to-end learning characteristic, deep neural networks that are used in computer vision research can be directly applied to audio signal processing area with \kg{minor} modifications. Since the magnitude spectrogram of \kg{an} audio \kg{signal} can be treated as a 2D single-channel image, convolutional neural networks (CNNs) have been \kg{successfully used in} various \kg{music applications, including the} source separation task~\cite{chandna2017monoaural,jansson2017singing}. While very deep CNNs are typically used in computer vision literature with very large datasets~\cite{szegedy2017inception,he2016deep}, CNNs used for audio source separation so far have relatively shallow architectures.

In this paper, we propose a novel music source separation framework using CNNs. We used stacked hourglass network~\cite{newell2016stacked} which \kg{was} originally proposed to solve human pose estimation in natural images. The CNNs take spectrogram images of \kg{a} music signal as inputs, and \nj{generate} masks for each music source to separate. An hourglass module captures both holistic features from low resolution feature maps and fine details from high resolution feature maps. The module outputs 3D volumetric data which has \nj{the} same width and height \nj{as} those of \nj{the} input spectrogram. \nj{The number} of \nj{output channels} equals the number of music sources to separate. The module is stacked for multiple times by taking the results of the previous module. As passing multiple modules, the results are refined and intermediate supervision helps faster learning in the initial state. We used a single network to separate multiple music sources, which \kg{reduces} \nj{both} time and space complexity for training \nj{as well as} testing.

We evaluated our framework on \nj{a couple of source separation tasks: 1) separating singing voice and accompaniments, and 2)} separating bass, drum, vocal, and other sounds from music. The results show that our method outperforms existing methods \nj{on} MIR-1K dataset~\cite{MIR_1K} and achieves competitive results \kg{comparable} to state-of-the-art methods \nj{on} DSD100 dataset~\cite{vincent2012signal} despite its simplicity.

\kg{The rest of the paper is organized as follows.} \sh{In Section~\ref{sec:rel}, we briefly review the literature of audio source separation focusing on DNN based methods. The proposed source separation framework and the architecture of the network are explained in Section~\ref{sec:method}. Experimental results are provided in Section~\ref{sec:exp}, and the paper is concluded in Section~\ref{sec:concl}.}

\section{Related Work}\label{sec:rel}

\begin{figure*}
 \centerline{
 \includegraphics[width=0.95\textwidth]{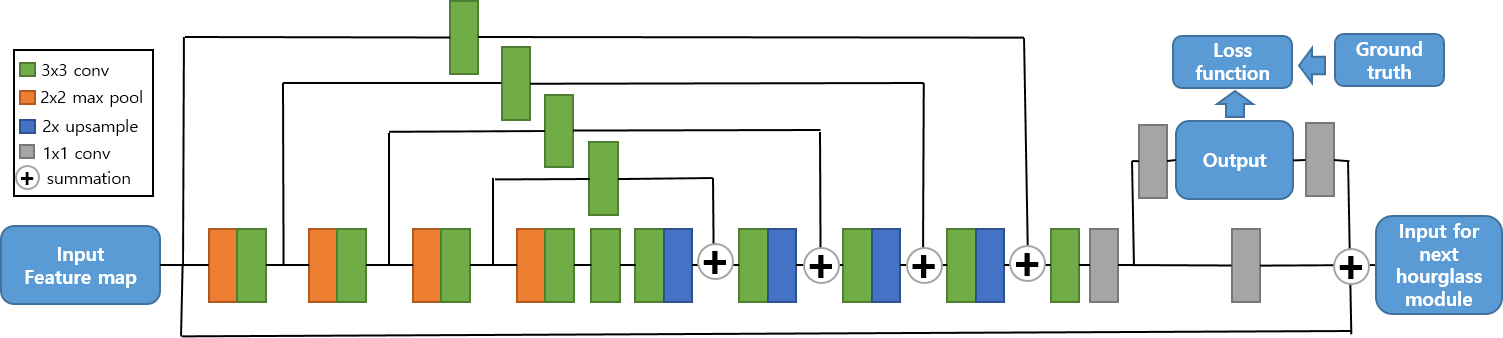}}
 \caption{Structure of the hourglass module used in this paper. We follow the structure proposed in~\cite{newell2017associative} except that the number of feature maps are set to 256 for all convolutional layers.}
 \label{fig_1}
\end{figure*}

Non-negative matrix factrization (NMF)~\cite{lee2001algorithms} is \nj{one of} the most widely-used \nj{algorithms} for audio source separation. \nj{It has been} successfully applied to monaural source separtion~\cite{virtanen2007monaural} and singing voice separation~\cite{zhang2015latent,vembu2005separation}. \nj{However, despite} its generality and flexibility, NMF is inferior to recently proposed \nj{DNN-based} methods in terms of performance and time complexity.

Simple deep feed-forward networks \nj{consisting} of multiple fully-connected layers showed reasonable performance for supervised audio source separation tasks~\cite{uhlich2015deep}. Wang \etal~\cite{wang2014training} used DNNs to learn \nj{an} ideal binary mask which boils the source separation \nj{problem} down to \nj{a} binary classification \nj{problem}. Simpson \etal~\cite{simpson2015deep} proposed \nj{a} convolutional DNN to predict \nj{a} probabilistic binary mask for singing voice separation. Recently, a fully complex-valued DNN~\cite{lee2017fully} is proposed to integrate phase information \nj{into the} magnitude spectrograms. \sh{Deep NMF~\cite{le2015deep} combined DNN and NMF by designing non-negative deep network and its back-propagation algorithm.}

Since \kg{an} audio signal is time series data, it is natural to use \kg{a sequence model like} recurrent neural networks (RNNs) for music source separation tasks to learn temporal information. Huang \etal~\cite{huang2015joint} proposed \nj{an} RNN framework that jointly \nj{optimizes} masks of foreground and background sources, \sh{which showed promising results for various source separation tasks. Other approaches include a recurrent encoder-decoder that exploits gated recurrent unit~\cite{Mimilakis2017} or discriminative RNN~\cite{wang2016discriminative}.}

CNNs are also an effective tool for audio signal analysis when the magnitude \nj{spectrogram is used} as an input. Fully convolutional networks (FCNs)~\cite{long2015fully} are initially proposed for semantic segmentation in the computer vision area, \nj{which} is also effective for solving human pose estimation~\cite{wei2016convolutional,newell2016stacked} or super-resolution~\cite{dong2016image}. FCNs usually contain downsampling and upsampling layers to learn meaningful features at multiple scales. Strided convolution or pooling is used for downsampling, while transposed convolution or nearest neighbor interpolation is mainly used for upsampling. It is proven that FCNs are also effective in signal processing. Chandna \etal~\cite{chandna2017monoaural} proposed encoder-decoder style FCN for monoaural audio source separation. Recently, singing voice separation using \nj{an} U-Net architecture~\cite{jansson2017singing} showed impressive performance. U-Net~\cite{ronneberger2015u} is a FCN which consists of a series of convolutional layers and upsampling layers. There is a skip connection which connects the convolutional layers of the same resolution. They trained vocal and accompaniment \nj{parts} separately on \nj{different networks}. Miron \etal~\cite{miron2017monaural} proposed the method that separates multiple sources using a single CNN. They used score-filtered spectrograms as inputs and generated masks for each source via \nj{an} encoder-decoder CNN. \sh{Multi-resolution FCN~\cite{grais2017multi} \kg{was} proposed for monaural audio source separation. Recently proposed CNN architecture~\cite{takahashi2017multi} based on DenseNet~\cite{iandola2014densenet} achieved state-of-the-art performance on DSD100 dataset.}

\section{Method}\label{sec:method}

\subsection{Network Architecture}

\begin{figure*}
 \centerline{
 \includegraphics[width=0.95\textwidth]{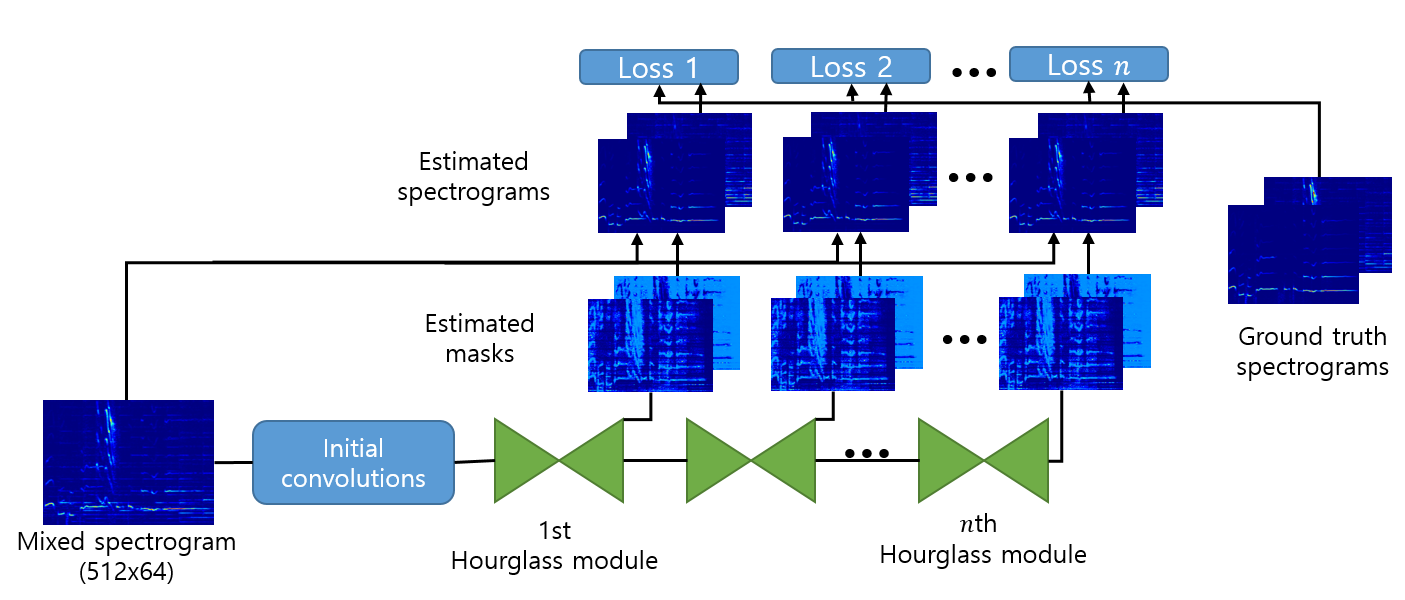}}
 \caption{Overall music source separation framework proposed in this paper. Multiple hourglass modules are stacked, and each module outputs masks for each music source. The masks are multiplied with the input spectrogram to generate predicted spectrograms. Differences between the estimated spectrograms and the ground truth ones are used as loss functions of the network.}
 \label{fig_2}
\end{figure*}

\nj{The stacked hourglass network}~\cite{newell2016stacked} was originally proposed to solve human pose estimation in RGB images. It is \nj{an} FCN \nj{consisting} of multiple hourglass modules. The hourglass module is similar to U-Net~\cite{ronneberger2015u}, \nj{of which feature} maps at lower \nj{(coarse)} resolution are obtained by repeatedly applying convolution and pooling operations. Then, the feature maps at \nj{the lowest} resolution are upsampled via nearest neighbor interpolation with a preceding convolutional layer. Feature maps at the same resolution in \nj{the} downsampling and \nj{the} upsampling \nj{steps} are connected with an additional convolutional layer. The hourglass module captures features at different scales by repeating pooling and upsampling with convolutional layers at each resolution. In addition, multiple hourglass modules are stacked to make the network deeper. As more hourglass modules are stacked, the network learns more powerful and informative features which \nj{refine} the estimation results. Loss functions are applied at the end of each module. This intermediate supervision improves training speed and performance of the network. 

\sh{The structure of a single hourglass module used in this paper is illustrated in Fig~\ref{fig_1}.} \shrev{Considering the efficiency and the size of the network}, we adopt the hourglass module used in~\cite{newell2017associative} which is a smaller network than \nj{the} originally proposed one \nj{in \cite{newell2016stacked}}. A notable difference is that the residual blocks~\cite{he2016deep} used in~\cite{newell2016stacked} are replaced with a single convolutional layer. \shrev{This light-weight structure showed competitive performance to the original network in human pose estimation with much smaller number of parameters.} In the module, there are four downsampling and upsampling steps. All convolutional layers in downsampling and upsampling steps have filter size of $3\times3$. \nj{The} $2\times2$ max pooling is used to halve the size of the feature maps, and \nj{the} nearest neighbor interpolation is used to double the size of the feature maps in \nj{the} upsampling steps. We fixed the \nj{size of the} maximum feature maps \nj{in} convolutional layers to 256 \nj{which is} different from~\cite{newell2017associative}. After the last upsampling layer, \nj{a} single $3\times3$ convolution and two $1\times1$ convolution is performed to generate network outputs. Then, \nj{an} $1\times1$ convolution is applied to the outputs to match the number of channels to that of the input feature maps. \nj{Another} $1\times1$ convolution is also applied to the feature maps which used for output generation. Finally, the two feature maps that passed \nj{the respective $1 \times 1$ convolution} and the input of the hourglass module is added together, and the resulting feature map is used as an input to the next hourglass module.

In the network used in this paper, input image firstly passes through initial convolutional layers that \nj{consist} of a $7\times7$ convolutional layer and four $3\times3$ convolutional layers where \nj{the} number of output feature maps for each layer is 64, 128, 128, 128, and 256 respectively. To make the output mask and the input spectrogram have the same size, we did not use the pooling operations in the initial convolutional layers before the hourglass module. The feature maps generated from the initial layers are fed to the first hourglass module. 
\nj{The proposed overall music source separation framework is depicted in Fig. \ref{fig_2}.}

\subsection{Music Source Separation}

\nj{As shown in Fig. \ref{fig_2}, to} apply the stacked hourglass network to music source separation, we aim to train the network to output soft masks for each music source given the magnitude spectrogram of \nj{the} mixed source. Hence, the output dimension of the network is $H \times W \times C$ where $H$ and $W$ are the height and width of the input spectrogram respectively, and $C$ is the number of music sources to separate. The magnitude spectrogram of separated music source is obtained by multiplying the mask and the input spectrogram. Our framework is scalable in that it requires almost no additional operation as \nj{the} number of sources \nj{increases}.

The input for the network is \nj{the} magnitude of spectrogram obtained from Short-Time Fourier Transform (STFT) with a \nj{window} size of 1024 and \nj{a} hop size of 256. The input source is downsampled to 8kHz to increase the duration of spectrograms in a batch and to speed up training. For each sample, magnitude spectrograms of mixed and separated \nj{sources} are generated, \nj{which} are divided by the maximum value of the mixed spectrogram for data normalization. The spectrograms have 512 frequency bins and the \nj{width of the spectrogram} depends on the duration of the music sources. \nj{For all the music sources, the width of the spectrogram is at least 64. Thus, 
we} fix the \nj{size of an input spectrogram} to $512\times64$. Hence, the size of the feature maps at the lowest resolution is $32 \times 4$. Starting time index is randomly chosen when the input batches are created.

Following ~\cite{ronneberger2015u}, we designed the loss function as an $L_{1,1}$ norm of the difference between the ground truth spectrogram and the estimated spectrogram. \nj{More concretely,} given \nj{an} input spectrogram $\mathbf{X}$, \nj{$i$th ground truth} music source $\mathbf{Y}_i$, and the generated mask for \nj{the} $i$th source in \nj{the} $j$th hourglass module $\mathbf{\hat{M}}_{ij}$, the loss for \nj{the} $i$th source is defined as
\begin{equation}
\mathcal{J}(i,j) = \| \mathbf{Y}_i - \mathbf{X} \odot \mathbf{\hat{M}}_{ij} \|_{1,1},
\end{equation}
where $\odot$ denotes element-wise multiplication of the matrix. $L_{1,1}$ norm is calculated as the sum of absolute values of matrix elements. The loss function of the network becomes
\begin{equation}
\mathcal{J} = \sum_{i=1}^{C} \sum_{j=1}^{D} \mathcal{J}(i,j),
\end{equation}
where $D$ is the number of hourglass modules stacked in the network. We directly used the output of the last $1 \times 1$ convolutional layer as the mask, which is different from~\cite{ronneberger2015u} \nj{where they} used \nj{the} sigmoid activation to generate masks. While it is natural to use \nj{the} sigmoid function to restrict the value of the mask to [0,1], we empirically found that not applying \nj{the} sigmoid function \nj{boosts} the training speed and improves the performance. Since sigmoid \nj{activations vanish} the gradient of the inputs that have large absolute values, \nj{they} may diminish the effect of intermediate supervision.

We \nj{have} stacked hourglass modules up to four and \nj{provide} analysis of the effect of stacking multiple modules in Section~\ref{sec:exp}. The network is trained using Adam optimizer~\cite{kingma2014adam} with \nj{a} starting learning rate of $10^{-4}$ and \nj{a} batch size of 4. 
\nj{We} trained the network for 15,000 and 150,000 iterations for MIR-1K dataset and DSD100 dataset respectively, and the learning rate is decreased to $2 \times 10^{-5}$ \nj{when} $80\%$ of the training is finished. \sh{No data augmentation is applied during training. The training took 3 hours for MIR-1K dataset and 31 hours for DSD100 dataset using a single GPU when the biggest model is used.} For the singing voice separation task, \nj{$C$} is set to 2 which corresponds to \textit{vocal} and \textit{accompaniments}. For the music source separation task in DSD100 dataset, \nj{$C=4$} is used where each output mask corresponds to \textit{drum}, \textit{bass}, \textit{vocal}, and \textit{others}. While it can be advantageous in terms of performance to train a network for a single source individually, it is computationally expensive to train a deep CNN for each source. Therefore, we trained a single network for each task.

In the test phase, the magnitude spectrogram of the input source is cropped to network input size and fed to the network sequentially. The output of the last hourglass module is used for testing. We set the negative values of output masks to 0 in order to avoid negative magnitude values. The masks are multiplied by the normalized magnitude spectrogram of the test source and unnormalized to generate spectrograms of separated sources. We did not change the phase spectrogram of the input source, and it is combined with the estimated magnitude spectrogram to retrieve signals for separated sources via inverse STFT.

\section{Experiments}\label{sec:exp}

We evaluated performance of the proposed method on MIR-1K and DSD100 datasets. For quantitative evaluation, we measured signal-to-distortion ratio (SDR), source-to-interference ratio (SIR), and source-to-artifacts ratio (SAR) based on BSS-EVAL metrics~\cite{vincent2006performance}. Normalized SDR (NSDR)~\cite{ozerov2007adaptation} is also measured for \nj{the} singing voice separation task which measures improvement between the mixture and the separated source. The values are obtained using mir-eval toolbox~\cite{raffel2014mir_eval}. Global NSDR (GNSDR), global SIR (GSIR), and global SAR (GSAR) are calculated as a weighted mean of NSDR, SIR, and SAR respectively whose weights are length of the source. The separated sources generated from the network are upsampled to the original sampling rate of the dataset and compared with ground truth sources for all experiments.

\subsection{MIR-1K dataset}

\begin{table}
 \begin{center}
 \begin{tabular}{|c|c|c|c|}
  \hline
  \multicolumn{4}{|c|}{Singing voice} \\
  \hline
  Method & GNSDR & GSIR & GSAR \\
  \hline
  MLRR~\cite{yang2013low}  & 3.85 & 5.63 & 10.70 \\
  DRNN~\cite{huang2015joint}  & 7.45 & 13.08 & 9.68 \\
  ModGD~\cite{sebastian2016group} & 7.50 & 13.73 & 9.45 \\
  U-Net~\cite{jansson2017singing} & 7.43 & 11.79 & 10.42 \\
  \hline
  SH-1stack  & 10.29 & 15.51 & 12.46 \\
  SH-2stack  & 10.45 & 15.89 & 12.49 \\
  SH-4stack  & \textbf{10.51} & \textbf{16.01} & \textbf{12.53} \\
  \hline
 \end{tabular}
 
 \vspace{1em} 
 
  \begin{tabular}{|c|c|c|c|}
  \hline
  \multicolumn{4}{|c|}{Accompaniments} \\
  \hline
  Method & GNSDR & GSIR & GSAR \\
  \hline
  MLRR~\cite{yang2013low}  & 4.19 & 7.80 & 8.22 \\
  U-Net~\cite{jansson2017singing} & 7.45 & 11.43 & 10.41 \\
  \hline
  SH-1stack  & 9.65 & 13.90 & 12.27 \\
  SH-2stack  & 9.64 & 13.69 & \textbf{12.39} \\
  SH-4stack  & \textbf{9.88} & \textbf{14.24} & 12.36 \\
  \hline
 \end{tabular}
\end{center}
 \caption{Quantitative evaluation of singing voice separation on MIR-1K dataset.}
 \label{tab_1}
\end{table}

\begin{figure*}
    \centering
    \begin{minipage}{0.32\textwidth}
        \centering
        \includegraphics[width=1\textwidth]{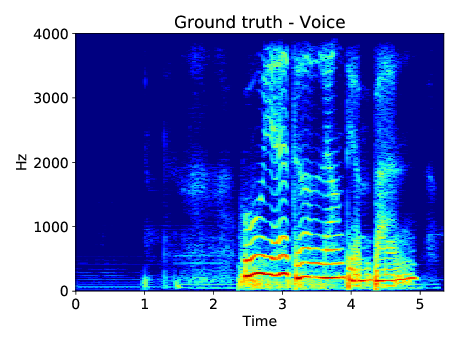}
        \label{fig_3_1}
    \end{minipage}%
    \begin{minipage}{0.32\textwidth}
        \centering
        \includegraphics[width=1\textwidth]{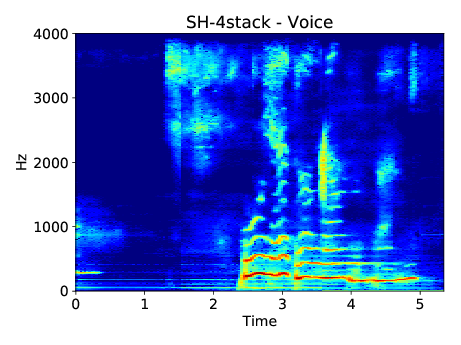}
        \label{fig3_2}
    \end{minipage}
     \begin{minipage}{0.32\textwidth}
        \centering
        \includegraphics[width=1\textwidth]{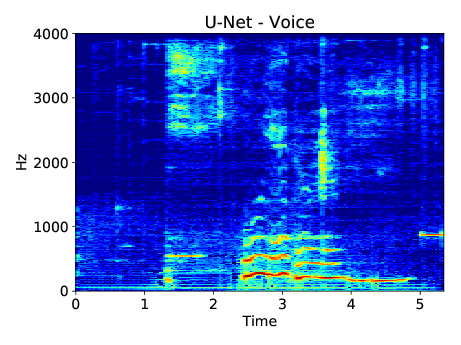}
        \label{fig3_3}
    \end{minipage}
    
    \vspace{-1em}
    
    \begin{minipage}{0.32\textwidth}
        \centering
        \includegraphics[width=1\textwidth]{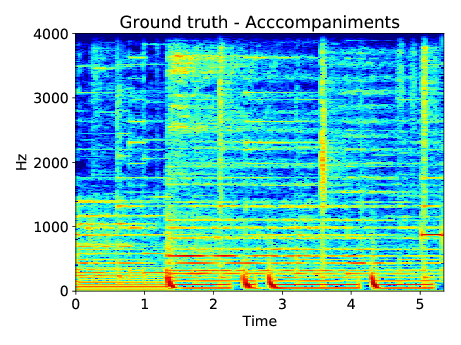}
        \label{fig_3_4}
    \end{minipage}%
    \begin{minipage}{0.32\textwidth}
        \centering
        \includegraphics[width=1\textwidth]{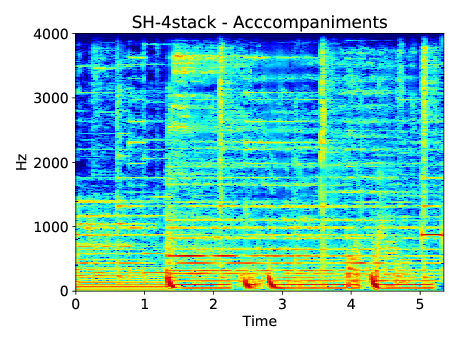}
        \label{fig3_5}
    \end{minipage}
     \begin{minipage}{0.32\textwidth}
        \centering
        \includegraphics[width=1\textwidth]{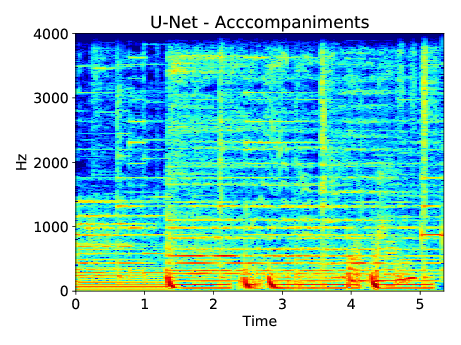}
        \label{fig3_6}
    \end{minipage}
\caption{Qualitative comparison of our method (SH-4stack) and U-Net for singing voice and accompaniments separation on \textit{annar\_3\_05} in MIR-1K dataset. Ground truth and estimated spectrograms are displayed in a log-scale. Our method is superior in capturing fine details compared to U-Net.}
\label{fig_3}
\end{figure*}

MIR-1K dataset is designed for singing voice separation research. It contains a thousand song clips extracted from 110 Chinese karaoke songs at a sampling rate of 16kHz. Following the previous works~\cite{huang2015joint,yang2013low}, we used one male and one female (\textit{abjones} and \textit{amy}) as a training set which \nj{contains} 175 clips in total. The remaining 825 clips are used for evaluation. For the baseline CNN, we trained the FCN that has U-Net~\cite{ronneberger2015u}-like structure and evaluated its performance. \shrev{We followed the structure of~\cite{jansson2017singing}, in which singing voice and accompaniments are trained on different networks.} For the stacked hourglass networks, both singing voice and accompaniments are obtained from a single network.

The evaluation results on test sets are shown in Table~\ref{tab_1}. We trained the networks with varying number of stacked hourglass modules 1, 2, and 4. It is proven that our stacked hourglass network (SH) significantly outperforms existing methods in all evaluation criteria. Our method gains 3.01 dB in GNSDR, 2.28 dB in GSIR, and 1.83 dB in GSAR compared to the best results of the existing methods. \shrev{It is also proven that the structure of the stacked hourglass module is more efficient and beneficial than U-Net~\cite{jansson2017singing} for music source separation. U-Net has 9.82 million parameters while single stack hourglass network has 8.99 million parameters considering only convolutional layers. Even with the absence of batch normalization, smaller number of parameters, and multi-source separation in a single network, the stacked hourglass network showed superior performance to U-Net.} While the network with a single hourglass module shows outstanding source separation performance, even better results are provided when multiple hourglass modules are stacked. This indicates \nj{that} SH network does not overfit \nj{even} when the network gets deeper despite small amount of the training data. Our method provides good performance on separating both singing voice and accompaniments with a single forward step.

Qualitative results of our method and comparison with U-Net are shown in Fig.~\ref{fig_3}. The estimated log spectrograms of singing voice and accompaniments from SH-4stack and U-Net and the ground truth log spectrograms are provided. It can be seen that our method captures fine details and harmonics compared to the U-Net. \shrev{The voice spectrogram from U-Net has more artifacts in the time slot of $0 \mathtt{\sim} 1$ and $4 \mathtt{\sim} 5$ compared to the result of SH-4stack. On the other hand, harmonics from voice signals can be clearly seen in the spectrogram of SH-4stack. For accompaniments spectrogram, it is observed that U-Net contains voice signals around the time slot of 3.}

\subsection{DSD100 dataset}

DSD100 dataset consists of 100 songs that are divided into 50 training sets and 50 test sets. For each song, four different music sources, bass, drums, vocals, and other as well as their mixtures are provided. The sources are stereophonic sound with a sampling rate of 44.1kHz. We converted all sources to monophonic and performed single channel source separation using stacked hourglass networks. We used a 4-stacked hourglass network (SH-4stack) for the experiments. 

The performance of music source separation using stacked hourglass network is provided in Table~\ref{tab_2}. We measured SDR of the separated sources for all test songs and report median values for comparison with existing methods. The methods that use single channel inputs are compared to our method. While the stacked hourglass network gives second-best performance following the state-of-the-art methods~\cite{takahashi2017multi} for drums and vocals, it shows poor performance for separating bass and other. This is mainly due to the similarity between bass and guitar sound in other sources, which confuses the network especially when trained together in a single network. Since the losses for all sources are summed up with equal weights, the network \nj{tends} to be trained to improve the separation performance of vocal and drum, which is easier than separating bass and other sources.

\begin{table}
\begin{center}
  \begin{tabular}{|c|c|c|c|c|}
  \hline
  Method & Bass & Drums & Other & Vocals \\
  \hline
  dNMF~\cite{weninger2014discriminative} & 0.91 & 1.87 & 2.43 & 2.56 \\
  DeepNMF~\cite{le2015deep} & 1.88 & 2.11 & 2.64 & 2.75 \\
  BLEND~\cite{uhlich2017improving}  &  2.76 & 3.93 & 3.37 & 5.13 \\
  MM-DenseNet~\cite{takahashi2017multi} & \textbf{3.91} & \textbf{5.37} & \textbf{3.81} & \textbf{6.00} \\
  \hline
  SH-4stack  & 1.77 & 4.11 & 2.36 & 5.16 \\
  \hline
 \end{tabular}
\end{center}
 \caption{Median SDR values for music source separation on DSD100 dataset.}
 \label{tab_2}
\end{table}

Next, we trained the stacked hourglass network for a singing voice separation task. The three sources except vocals are mixed together to form accompaniments source. The median SDR values for each source are reported in Table~\ref{tab_3}. Our method achieved best result for accompaniments separation and second-best for vocal separation. Separation performance of vocals is improved compared to the music source separation setting. It can be inferred that the stacked hourglass network provides better results as number of sources are smaller and the separating sources are more distinguishable from each other.

\begin{table}
\begin{center}
  \begin{tabular}{|c|c|c|}
  \hline
  Method & \;\;Vocals\;\; & Accompaniments \\
  \hline
  DeepNMF~\cite{le2015deep} & 2.75 & 8.90 \\
  wRPCA~\cite{jeong2017singing} & 3.92 & 9.45 \\
  NUG~\cite{nugraha2016multichannel}  &  4.55  & 10.29 \\
  BLEND~\cite{uhlich2017improving}  &  5.23  & 11.70 \\
  MM-DenseNet~\cite{takahashi2017multi} & \textbf{6.00} & 12.10 \\
  \hline
  SH-4stack  & 5.45 & \textbf{12.14} \\
  \hline
 \end{tabular}
\end{center}
 \caption{Median SDR values for singing voice separation on DSD100 dataset.}
 \label{tab_3}
\end{table}

\begin{figure*}[t]
    \centering
    \begin{minipage}{0.45\textwidth}
        \centering        \includegraphics[width=1\textwidth]{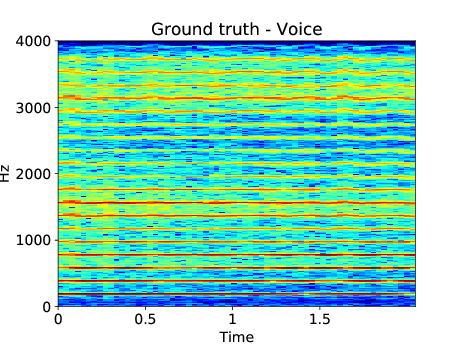}
        \label{fig_4_1}
    \end{minipage}
    \begin{minipage}{0.45\textwidth}
        \centering        \includegraphics[width=1\textwidth]{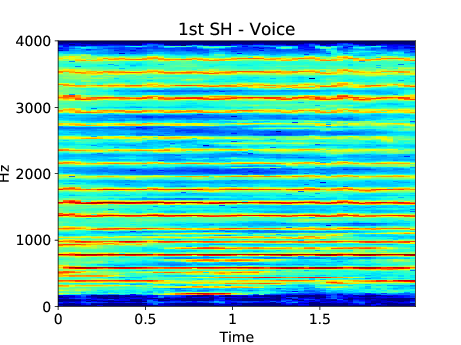}
        \label{fig_4_2}
    \end{minipage}
    \begin{minipage}{0.45\textwidth}
        \centering        \includegraphics[width=1\textwidth]{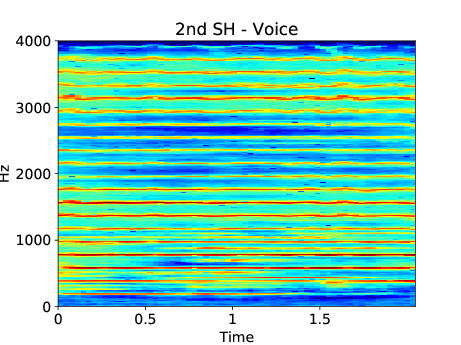}
        \label{fig_4_3}
    \end{minipage}
    \begin{minipage}{0.45\textwidth}
        \centering        \includegraphics[width=1\textwidth]{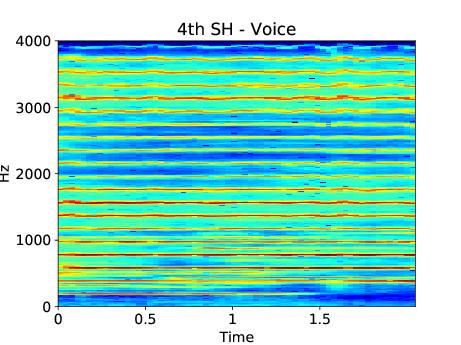}
        \label{fig_4_4}
    \end{minipage}
\caption{Examples showing the effectiveness of stacking multiple hourglass modules. Ground truth and estimated spectrograms of the part of the song \textit{Schoolboy Fascination} in DSD100 dataset are shown. SDR values of the source generated from the spectrograms obtained from first, second, fourth hourglass module are \shrev{10.90, 12.50, 13.30 respectively. Especially, it is observed that the estimated spectrogram captures fine details of spectrogram at low frequency range ($0 \mathtt{\sim} 500$ Hz) as more hourglass modules are stacked.}}
\label{fig_4}
\end{figure*}

Lastly, we investigate how the stacked hourglass network improves the output masks as they pass through the hourglass modules within the network. The example illustrated in Fig.~\ref{fig_4} shows the estimated \nj{voice} spectrogram of first, second, and fourth hourglass module with the ground truth spectrogram from one of the test sets of DSD 100 dataset. \shrev{It is observed that the estimated spectrogram becomes more similar to the ground truth as it is generated from a deeper side of the network. In the result of the fourth hourglass module, spectrograms at low frequency are clearly recovered compared to the result of the first hourglass module. The artifacts in the range of $2000 \mathtt{\sim} 3000$ Hz are also removed.} Although it is hard to recognize the difference in the spectrogram image, the difference of SDR between the source estimated from the first hourglass module and the last hourglass module is about 2.4dB which is a significant performance gain.

\section{Conclusion}\label{sec:concl}

In this paper, we proposed music source separation algorithm using stacked hourglass networks. The network successfully captures features at both coarse and fine resolution, and it produces masks that are applied to the input spectrograms. Multiple hourglass modules refines the estimation results and outputs the better results. Experimental results has proven the effectiveness of the proposed framework for music source separation. We implemented the framework in its simplest form, and there is a lot of room \nj{for performance improvements} including data augmentation, regularization of CNNs, and ensemble learning of multiple models. Designing a loss function that considers correlation of different sources may further improves the performance.

\section{Acknowledgement}
This work was supported by Next-Generation Information Computing Development Program through the National Research Foundation of Korea (2017M3C4A7077582).

\bibliography{ISMIRtemplate}

\begin{thebibliography}{10}

\bibitem{chandna2017monoaural}
Pritish Chandna, Marius Miron, Jordi Janer, and Emilia G{\'o}mez.
\newblock Monoaural audio source separation using deep convolutional neural
  networks.
\newblock In {\em International Conference on Latent Variable Analysis and
  Signal Separation}, pages 258--266. Springer, 2017.

\bibitem{dong2016image}
Chao Dong, Chen~Change Loy, Kaiming He, and Xiaoou Tang.
\newblock Image super-resolution using deep convolutional networks.
\newblock {\em IEEE transactions on pattern analysis and machine intelligence},
  38(2):295--307, 2016.

\bibitem{grais2017multi}
Emad~M Grais, Hagen Wierstorf, Dominic Ward, and Mark~D Plumbley.
\newblock Multi-resolution fully convolutional neural networks for monaural
  audio source separation.
\newblock {\em arXiv preprint arXiv:1710.11473}, 2017.

\bibitem{he2016deep}
Kaiming He, Xiangyu Zhang, Shaoqing Ren, and Jian Sun.
\newblock Deep residual learning for image recognition.
\newblock In {\em Proceedings of the IEEE conference on computer vision and
  pattern recognition}, pages 770--778, 2016.

\bibitem{MIR_1K}
C.~L. Hsu and J.~S.~R. Jang.
\newblock On the improvement of singing voice separation for monaural
  recordings using the mir-1k dataset.
\newblock {\em IEEE Transactions on Audio, Speech, and Language Processing},
  18(2):310--319, Feb 2010.

\bibitem{huang2015joint}
Po-Sen Huang, Minje Kim, Mark Hasegawa-Johnson, and Paris Smaragdis.
\newblock Joint optimization of masks and deep recurrent neural networks for
  monaural source separation.
\newblock {\em IEEE/ACM Transactions on Audio, Speech, and Language
  Processing}, 23(12):2136--2147, 2015.

\bibitem{iandola2014densenet}
Forrest Iandola, Matt Moskewicz, Sergey Karayev, Ross Girshick, Trevor Darrell,
  and Kurt Keutzer.
\newblock Densenet: Implementing efficient convnet descriptor pyramids.
\newblock {\em arXiv preprint arXiv:1404.1869}, 2014.

\bibitem{jansson2017singing}
Andreas Jansson, Eric Humphrey, Nicola Montecchio, Rachel Bittner, Aparna
  Kumar, and Tillman Weyde.
\newblock Singing voice separation with deep u-net convolutional networks.
\newblock {\em 18th International Society for Music Information Retrieval
  Conferenceng, Suzhou, China}, 2017.

\bibitem{jeong2017singing}
Il-Young Jeong and Kyogu Lee.
\newblock Singing voice separation using rpca with weighted $ $ l\_ $\{$1$\}$ $
  $-norm.
\newblock In {\em International Conference on Latent Variable Analysis and
  Signal Separation}, pages 553--562. Springer, 2017.

\bibitem{kingma2014adam}
Diederik~P Kingma and Jimmy Ba.
\newblock Adam: A method for stochastic optimization.
\newblock {\em arXiv preprint arXiv:1412.6980}, 2014.

\bibitem{le2015deep}
Jonathan Le~Roux, John~R Hershey, and Felix Weninger.
\newblock Deep nmf for speech separation.
\newblock In {\em Acoustics, Speech and Signal Processing (ICASSP), 2015 IEEE
  International Conference on}, pages 66--70. IEEE, 2015.

\bibitem{lee2001algorithms}
Daniel~D Lee and H~Sebastian Seung.
\newblock Algorithms for non-negative matrix factorization.
\newblock In {\em Advances in neural information processing systems}, pages
  556--562, 2001.

\bibitem{lee2017fully}
Yuan-Shan Lee, Chien-Yao Wang, Shu-Fan Wang, Jia-Ching Wang, and Chung-Hsien
  Wu.
\newblock Fully complex deep neural network for phase-incorporating monaural
  source separation.
\newblock In {\em Acoustics, Speech and Signal Processing (ICASSP), 2017 IEEE
  International Conference on}, pages 281--285. IEEE, 2017.

\bibitem{long2015fully}
Jonathan Long, Evan Shelhamer, and Trevor Darrell.
\newblock Fully convolutional networks for semantic segmentation.
\newblock In {\em Proceedings of the IEEE conference on computer vision and
  pattern recognition}, pages 3431--3440, 2015.

\bibitem{Mimilakis2017}
Stylianos~Ioannis Mimilakis, Konstantinos Drossos, Tuomas Virtanen, and Gerald
  Schuller.
\newblock A recurrent encoder-decoder approach with skip-filtering connections
  for monaural singing voice separation.
\newblock {\em CoRR}, abs/1709.00611, 2017.

\bibitem{miron2017monaural}
Marius Miron, Jordi Janer, and Emilia G{\'o}mez.
\newblock Monaural score-informed source separation for classical music using
  convolutional neural networks.
\newblock In {\em 18th International Society for Music Information Retrieval
  Conference, Suzhou, China}, 2017.

\bibitem{newell2017associative}
Alejandro Newell, Zhiao Huang, and Jia Deng.
\newblock Associative embedding: End-to-end learning for joint detection and
  grouping.
\newblock In {\em Advances in Neural Information Processing Systems}, pages
  2274--2284, 2017.

\bibitem{newell2016stacked}
Alejandro Newell, Kaiyu Yang, and Jia Deng.
\newblock Stacked hourglass networks for human pose estimation.
\newblock In {\em European Conference on Computer Vision}, pages 483--499.
  Springer, 2016.

\bibitem{nugraha2016multichannel}
Aditya~Arie Nugraha, Antoine Liutkus, and Emmanuel Vincent.
\newblock Multichannel music separation with deep neural networks.
\newblock In {\em Signal Processing Conference (EUSIPCO), 2016 24th European},
  pages 1748--1752. IEEE, 2016.

\bibitem{ozerov2007adaptation}
Alexey Ozerov, Pierrick Philippe, Frdric Bimbot, and Rmi Gribonval.
\newblock Adaptation of bayesian models for single-channel source separation
  and its application to voice/music separation in popular songs.
\newblock {\em IEEE Transactions on Audio, Speech, and Language Processing},
  15(5):1564--1578, 2007.

\bibitem{raffel2014mir_eval}
Colin Raffel, Brian McFee, Eric~J Humphrey, Justin Salamon, Oriol Nieto, Dawen
  Liang, Daniel~PW Ellis, and C~Colin Raffel.
\newblock mir\_eval: A transparent implementation of common mir metrics.
\newblock In {\em In Proceedings of the 15th International Society for Music
  Information Retrieval Conference, ISMIR}. Citeseer, 2014.

\bibitem{ronneberger2015u}
Olaf Ronneberger, Philipp Fischer, and Thomas Brox.
\newblock U-net: Convolutional networks for biomedical image segmentation.
\newblock In {\em International Conference on Medical image computing and
  computer-assisted intervention}, pages 234--241. Springer, 2015.

\bibitem{sebastian2016group}
Jilt Sebastian and Hema~A Murthy.
\newblock Group delay based music source separation using deep recurrent neural
  networks.
\newblock In {\em Signal Processing and Communications (SPCOM), 2016
  International Conference on}, pages 1--5. IEEE, 2016.

\bibitem{simpson2015deep}
Andrew~JR Simpson, Gerard Roma, and Mark~D Plumbley.
\newblock Deep karaoke: Extracting vocals from musical mixtures using a
  convolutional deep neural network.
\newblock In {\em International Conference on Latent Variable Analysis and
  Signal Separation}, pages 429--436. Springer, 2015.

\bibitem{szegedy2017inception}
Christian Szegedy, Sergey Ioffe, Vincent Vanhoucke, and Alexander~A Alemi.
\newblock Inception-v4, inception-resnet and the impact of residual connections
  on learning.
\newblock In {\em AAAI}, volume~4, page~12, 2017.

\bibitem{takahashi2017multi}
Naoya Takahashi and Yuki Mitsufuji.
\newblock Multi-scale multi-band densenets for audio source separation.
\newblock In {\em Applications of Signal Processing to Audio and Acoustics
  (WASPAA), 2017 IEEE Workshop on}, pages 21--25. IEEE, 2017.

\bibitem{uhlich2015deep}
Stefan Uhlich, Franck Giron, and Yuki Mitsufuji.
\newblock Deep neural network based instrument extraction from music.
\newblock In {\em Acoustics, Speech and Signal Processing (ICASSP), 2015 IEEE
  International Conference on}, pages 2135--2139. IEEE, 2015.

\bibitem{uhlich2017improving}
Stefan Uhlich, Marcello Porcu, Franck Giron, Michael Enenkl, Thomas Kemp, Naoya
  Takahashi, and Yuki Mitsufuji.
\newblock Improving music source separation based on deep neural networks
  through data augmentation and network blending.
\newblock In {\em Acoustics, Speech and Signal Processing (ICASSP), 2017 IEEE
  International Conference on}, pages 261--265. IEEE, 2017.

\bibitem{vembu2005separation}
Shankar Vembu and Stephan Baumann.
\newblock Separation of vocals from polyphonic audio recordings.
\newblock In {\em ISMIR}, pages 337--344. Citeseer, 2005.

\bibitem{vincent2012signal}
Emmanuel Vincent, Shoko Araki, Fabian Theis, Guido Nolte, Pau Bofill, Hiroshi
  Sawada, Alexey Ozerov, Vikrham Gowreesunker, Dominik Lutter, and Ngoc~QK
  Duong.
\newblock The signal separation evaluation campaign (2007--2010): Achievements
  and remaining challenges.
\newblock {\em Signal Processing}, 92(8):1928--1936, 2012.

\bibitem{vincent2006performance}
Emmanuel Vincent, R{\'e}mi Gribonval, and C{\'e}dric F{\'e}votte.
\newblock Performance measurement in blind audio source separation.
\newblock {\em IEEE transactions on audio, speech, and language processing},
  14(4):1462--1469, 2006.

\bibitem{virtanen2007monaural}
Tuomas Virtanen.
\newblock Monaural sound source separation by nonnegative matrix factorization
  with temporal continuity and sparseness criteria.
\newblock {\em IEEE transactions on audio, speech, and language processing},
  15(3):1066--1074, 2007.

\bibitem{wang2016discriminative}
Guan-Xiang Wang, Chung-Chien Hsu, and Jen-Tzung Chien.
\newblock Discriminative deep recurrent neural networks for monaural speech
  separation.
\newblock In {\em Acoustics, Speech and Signal Processing (ICASSP), 2016 IEEE
  International Conference on}, pages 2544--2548. IEEE, 2016.

\bibitem{wang2014training}
Yuxuan Wang, Arun Narayanan, and DeLiang Wang.
\newblock On training targets for supervised speech separation.
\newblock {\em IEEE/ACM Transactions on Audio, Speech and Language Processing
  (TASLP)}, 22(12):1849--1858, 2014.

\bibitem{wei2016convolutional}
Shih-En Wei, Varun Ramakrishna, Takeo Kanade, and Yaser Sheikh.
\newblock Convolutional pose machines.
\newblock In {\em Proceedings of the IEEE Conference on Computer Vision and
  Pattern Recognition}, pages 4724--4732, 2016.

\bibitem{weninger2014discriminative}
Felix Weninger, Jonathan~Le Roux, John~R Hershey, and Shinji Watanabe.
\newblock Discriminative nmf and its application to single-channel source
  separation.
\newblock In {\em Fifteenth Annual Conference of the International Speech
  Communication Association}, 2014.

\bibitem{yang2013low}
Yi-Hsuan Yang.
\newblock Low-rank representation of both singing voice and music accompaniment
  via learned dictionaries.
\newblock In {\em ISMIR}, pages 427--432, 2013.

\bibitem{zhang2015latent}
Xiu Zhang, Wei Li, and Bilei Zhu.
\newblock Latent time-frequency component analysis: A novel pitch-based
  approach for singing voice separation.
\newblock In {\em Acoustics, Speech and Signal Processing (ICASSP), 2015 IEEE
  International Conference on}, pages 131--135. IEEE, 2015.

\end{thebibliography}

\end{document}